\documentclass[  
aps,
pra,
amsmath,amssymb,showpacs,
reprint,
floatfix,
nofootinbib,
longbibliography,
]{revtex4-2}

\usepackage{amsmath,amsfonts,amssymb}
\usepackage{graphicx}

\newcommand{\est}{\varphi_{\textrm{est}}}

\allowdisplaybreaks[3]

\graphicspath{{./fig}}

\begin{document}
	
	\title{Interferometric characterization of the relative phase between two X-ray free-electron laser pulses using long-lived M\"ossbauer resonances}
	
	\author{Lukas \surname{Wolff}}
	\affiliation{Max-Planck-Institut f\"ur Kernphysik, Saupfercheckweg 1, 69117 Heidelberg, Germany}
	
	\author{J\"org \surname{Evers}}
	\affiliation{Max-Planck-Institut f\"ur Kernphysik, Saupfercheckweg 1, 69117 Heidelberg, Germany}
	
	\begin{abstract}
        Coherence-based spectroscopy methods are powerful tools to explore structure and dynamics of matter. However, towards higher photon energies, the generation of sequences of pulses with well-characterized relative delays and phases remains a challenge. Here, we introduce a method to measure the relative phase $\varphi$ between subsequent transform-limited pulses from high-repetition-rate x-ray free-electron lasers (XFELs). It is based on a Ramsey-type interference measurement, enabled by introducing  long-lived M\"ossbauer resonances into the XFEL beam path up- or downstream a primary experiment, which allow one to bridge the temporal gap between the XFEL pulses. The measured phase can be used as additional input for the analysis of the primary experiment. 
	\end{abstract}
	
	\maketitle

In recent years, x-ray free-electron lasers (XFELs) have arisen as powerful tools complementing existing synchrotron light sources and high-harmonic generation  from the extreme ultraviolet to the hard x-ray regime. Key features are the high brilliance, ultrashort pulses down to the femto- and even attosecond scale and the extraordinary  coherence properties of the XFEL pulses~\cite{jaeschke2016}. These features render them ideal candidates, e.g., for studying dynamics on small length scales and site-specific phenomena on fast and ultrafast time scales. 

In the longer-wavelength part of the electromagnetic spectrum, coherent nonlinear spectroscopic techniques such as coherent pump-probe, photon-echo and coherent multidimensional spectroscopy~\cite{Aue1976,mukamel,Hamm2011,Cho2008} provide unique insights into coherent quantum dynamics of molecular and atomic systems~\cite{Zewail2000}. They rely on interferometric measurements enabled by sequences of short pulses with well-defined relative timings and phases.
Extending such coherence-based schemes to higher photon energies is highly desirable, for example, since it promises site-specific addressing  via localized inner-core resonances and high time resolution~\cite{Young_2018,Ramasesha,Kraus2018,chergui2023}. However, it remains challenging to meet the extreme stability requirements on the pulses. Two complementary experimental platforms for this are high-harmonic generation (HHG)~\cite{Agostini2004,RevModPhys.81.163,corkum2007,Protopapas1997} and XFELs. HHG typically provides superior coherence properties as compared to XFELs, but achieves lower pulse intensities  and is limited in photon energy. The XFEL coherence can be improved via seeding~\cite{amann2012,inoue2019,nam2021,liu2023}. In particular, the Free Electron Laser Radiation for Multidisciplinary Investigations (FERMI) light source employs external laser seeding to provide fully coherent XUV or soft x-ray light, and has been used to demonstrate the shaping of the temporal pulse phase~\cite{PhysRevLett.115.114801} and coherent control~\cite{Prince2016} in the XUV domain. It also allows to control the relative phase and timing properties separately~\cite{wituschek2020}. 

These developments focus on fast dynamics on the femtosecond time scale.  However, slower dynamics also is of considerable interest. For instance, x-ray photon correlation spectroscopy (XPCS)~\cite{Madsen2020} and X-ray speckle visibility spectroscopy (XSVS)~\cite{Lehmkuehler2021} are employed very successfully to study dynamics over a vast range of time scales, from femtoseconds to hours. 
However,  from the nano- to the micro- or milli-second range, a temporal gap for studying the dynamics of complex matter presently exists~\cite{Lehmkuehler2021,schroer2018petra} which is difficult to access experimentally. Diffraction-limited storage rings~\cite{Eriksson:vv0002,doi:10.1080/08940886.2016.1244462,schroer2018petra,Weckert:it5005}, multi-bucket XFEL pulses~\cite{decker2022} and high repetition-rate superconducting XFELs~\cite{ROSSBACH2019,decking2020,galayda2018,zhu2017} are expected to close this gap.One example for slower dynamics is nuclear quantum optics~\cite{Adams2013,Adams2019,Kuznetsova2017,rohlsberger2020,Roehlsberger2021}, typically based on the narrow resonances in the M\"ossbauer isotope ${}^{57}$Fe at 14.4~keV transition energy, which operates on time scales of the order of the nuclear lifetime  $\tau = 141$~ns~\cite{RoehlsbergerBook}. Like in the XUV, spectral pulse shaping~\cite{Heeg2017} and coherent control of nuclear dynamics~\cite{Heeg2021} have already been demonstrated, as well as other coherent control schemes based on dynamical control of the magnetization~\cite{Shvydko1996,PhysRevLett.103.017401,PhysRevLett.109.197403}, magnon excitation~\cite{bocklage2021}, sample motion~\cite{Helisto1991,Schindelmann2002,Vagizov2014,Heeg2017,Heeg2021,velten2024}, or suitably tailored photonic environments~\cite{Roehlsberger2010,Roehlsberger2012EIT,Heeg2013,Heeg2013SGC,Heeg2015SL,haber_collective_2016,Haber2017,Lentrodt2020}. However, so far, a source is lacking which is capable of delivering coherent sequences of short and temporally separated pulses with well-defined phase relations.

This raises the question whether it is possible to coherently study and control dynamics in the  nanosecond to microsecond regime using the inherent pulse structure of seeded hard x-ray superconducting XFELs. One approach is to measure the relative phase between  transform-limited XFEL pulses, thereby gaining additional information for the primary experiment.

Here, we introduce a method to characterize this relative phase $\varphi$ between transform-limited XFEL pulses. To this end, a target containing nuclei featuring a long-lived M\"ossbauer transition is added to the x-ray beam path up- or downstream of the primary experiment. The nuclear resonant scattering provides a slow time scale to the experiment, which allows one to bridge the temporal gap between two consecutive x-ray pulses. This way, a Ramsey-type interferometric measurement of the pulse-to-pulse relative phase $\varphi$ becomes possible under state-of-the-art experimental conditions. The resulting phase information can then be used to analyze the measurement data of  the  primary experiment. For the future, one may also envision a stabilization of the XFEL pulse-to-pulse phase based on the phase measurement as a feedback to the x-ray source.

\begin{figure}[t]
\includegraphics[width = \columnwidth]{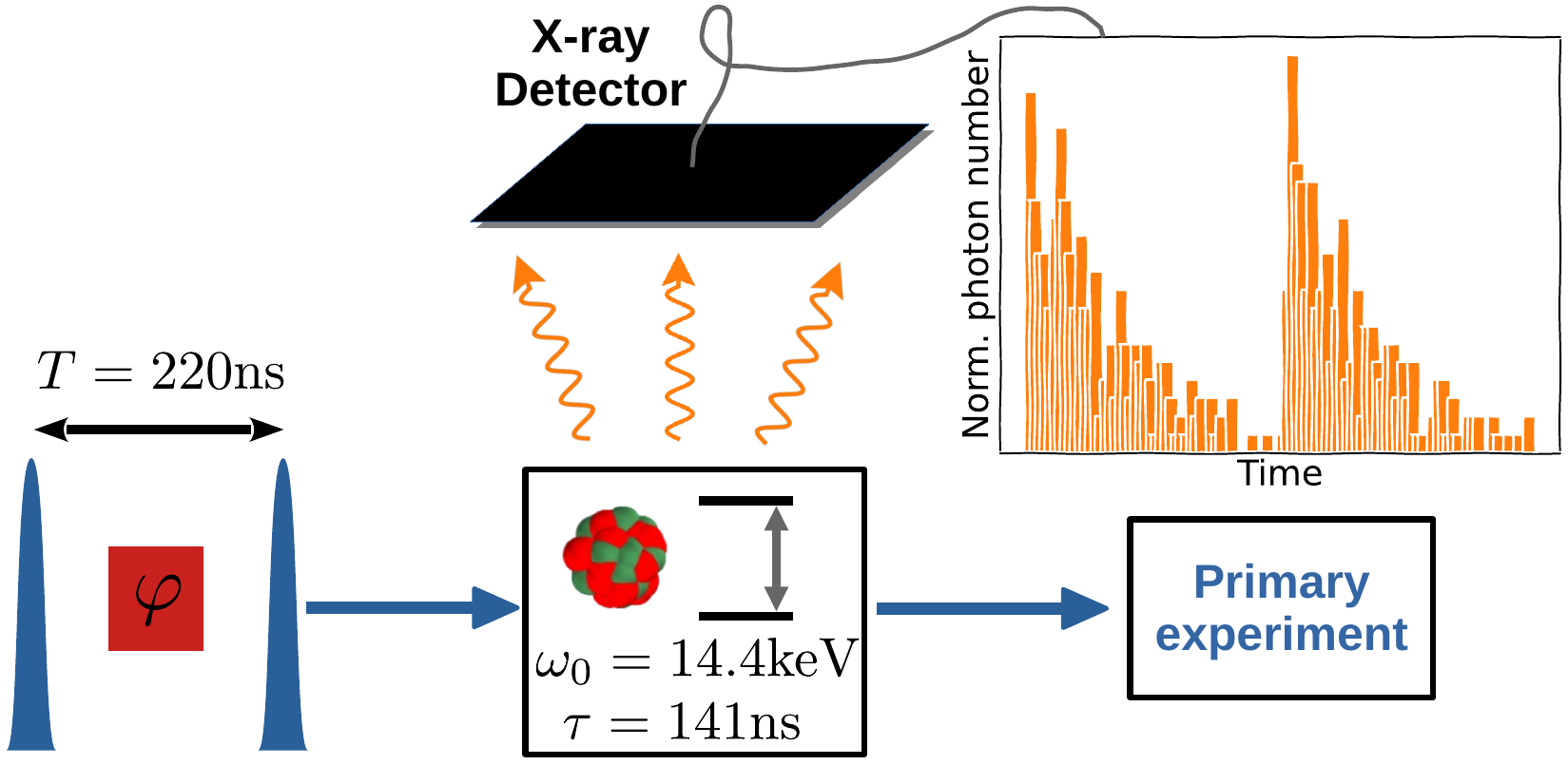}
\caption{Schematic setup for the characterization of the relative phase $\varphi$ between two transform-limited XFEL pulses with pulse separation $T$. A target containing  M{\"o}ssbauer nuclei with resonance frequency $\omega_0$ and lifetime $\tau$ is placed in the beam path.  The intensity of incoherent nuclear scattering of the two x-ray pulses  is measured as function of time. Since  the nuclear lifetime $\tau$ is comparable to $T$, the intensity  contains an interference contribution depending on $\varphi$. Therefore, $\varphi$ can be reconstructed and utilized as additional information for  the  primary experiment. \label{fig-Main}}
\end{figure}

A schematic setup of the phase characterization method is shown in Fig.~\ref{fig-Main}. The goal is to measure the relative phase $\varphi$ of two consecutive XFEL pulses with temporal separation $T$. A target containing M{\"o}ssbauer nuclei is placed in the beam path of an XFEL, and the resonantly scattered radiation is detected via spatially incoherent nuclear resonant scattering. The observation of spatially incoherent scattering emitted into $4\pi$~\cite{Smirnov2007, SmirnovKohn1995, Bergmann1994,sturhahn1999} allows one to operate the phase measurement without significantly perturbing the incident beam before a subsequent downstream primary experiment. 

There are two main requirements for gaining phase information. First, the nuclear lifetime has to be comparable to the pulse separation $T$. Only then, the nuclear scattering allows for an interferometric measurement involving two consecutive XFEL pulses. Correspondingly, the long-lived nuclear resonance only overlaps with a tiny part  of the incident pulse spectrum, such that primary experiments operating with broader resonances  are largely unaffected by the nuclei. Second, the nuclear-resonant flux has to be sufficiently high to allow for a measurement of the interferometric signal for individual x-ray pulse pairs. 

Regarding the first point, we consider the 4.5 MHz mode of EuXFEL, with intra-pulse separations $T \approx 220$~ns~\cite{Tschentscher2017}. This time scale is compatible with the natural lifetime $\tau = 141$~ns of the archetype $\hbar \omega_0 = 14.4$keV transition of the M{\"o}ssbauer isotope ${}^{57}\textrm{Fe}$. Correspondingly, the natural line-width $\hbar \gamma = 4.7$~neV is much smaller than the $\sim$eV pulse bandwidth in hard x-ray self-seeding operation~\cite{amann2012,inoue2019,nam2021,liu2023}, such that the overall effect on the short x-ray pulse is negligible for the primary experiment. Note that our approach therefore a priori characterizes the relative phase of a particular frequency component of the pulses. However, approaching transform-limited x-ray pulses with flat spectral phase profile via seeding~\cite{amann2012,inoue2019,nam2021,liu2023} enables one to extrapolate this phase information across the entire pulse. 
Regarding the second point, the first XFEL experiment probing ${}^{57}$Fe M{\"o}ssbauer nuclei has already demonstrated the possibility to record few tens of signal photons per exciting x-ray pulse~\cite{Chumakov2018}. An experiment at EuXFEL in hard x-ray self-seeding mode has succeeded in resonantly exciting the M\"ossbauer isotope ${}^{45}$Sc which has a single-nucleus line-width in the feV range~\cite{shvyd2023}. Recent experiments at EuXFEL have further observed several hundreds of signal photons coherently scattered off ${}^{57}$Fe per x-ray excitation~\cite{EuXFEL57}, and we will show below that these conditions are sufficient for gaining information on the relative phase $\varphi$.

To analyze the phase recovery, we model the incident field of two consecutive seeded XFEL pulses separated by time $T$ acting on the nuclei as 
\begin{align}
E_{\textrm{in}}(t,T,\varphi) = E_0 \,e^{-i\nu t}\:\left[\delta(t) + e^{i\varphi}\delta(t-T)\right]\,.  \label{eq-EIn}
\end{align}
Here, the x-ray field amplitude is $E_0$, and $\nu$ the carrier frequency. Since the  time scales of the nuclear dynamics are orders of magnitude slower than the x-ray pulse duration, the latter are approximated as $\delta(t)$-flashes. 

To obtain the spatially incoherent scattering emitted into the full solid angle of $4\pi$, one first considers the light resonantly scattered off a single infinitesimal slice of the nuclear target at depth $z$ on resonance~\cite{Smirnov1999,Smirnov2007,HannonTrammell1999},
\begin{align}
I_{4\pi}(t, T, \varphi, z) &= I_0 \Theta(t)\bigl [  \mathcal{E}^2_{4\pi}(t,z) + \mathcal{E}^2_{4\pi}(t-T,z) \Theta(t-T)  \nonumber\\[1ex]
  + 2\cos &(\varphi)\, \mathcal{E}_{4\pi}(t,z) \mathcal{E}_{4\pi}(t-T,z) \Theta(t-T) \bigr ] \, , \label{eq-TwoLevelDist} \\[1ex]
\mathcal{E}_{4\pi}(t,z) &= e^{-\frac{\mu_e}{2}z}e^{-\frac{\gamma}{2}t}J_0(2\sqrt{b(z)t})  \, . \label{eq:response}
\end{align}
Here, $\mu_e$ denotes the electronic absorption length, $\gamma = \tau^{-1}$ the single-nucleus line-width, $\omega_0$ the nuclear resonance frequency, $b(z)$ the target thickness parameter up to depth $z$, $J_0$ the zeroth-order Bessel function of the first kind, and $\Theta(t)$ the Heaviside step function.
$\mathcal{E}_{4\pi}(z,t)$ is the real-valued envelope function of the induced nuclear dipole moment~\cite{Smirnov2007,SmirnovKohn1995,Smirnov1999}. For simplicity, Eq.~(\ref{eq-TwoLevelDist}) assumes resonant interaction $\nu = \omega_0$, and a  nuclear target without hyperfine splitting. The total incoherently emitted intensity  $I_{\textrm{tot}}(t,T,\varphi)$ is then obtained by integrating Eq.~(\ref{eq-TwoLevelDist}) over the target thickness from $z = 0$ to $z = z_{\textrm{max}}$.

From Eq.~(\ref{eq-TwoLevelDist}) we find that the incoherently scattered intensity contains an interference contribution which depends on the pulse-to-pulse phase $\varphi$.  In the following, we show that it allows one to reconstruct $\varphi$ from measured data. In Appendix~\ref{app:detuning} we further show that the qualitative structure of Eq.~(\ref{eq-TwoLevelDist}) with its $\varphi$-dependence persists for non-zero detuning $\nu \neq \omega_0$, in the case of nuclear forward scattering~\cite{Smirnov1999,HannonTrammell1999,RoehlsbergerBook}, and in the presence of hyperfine splitting. Therefore, our results qualitatively generalize to these settings.

\begin{figure}
\includegraphics[width = \columnwidth]{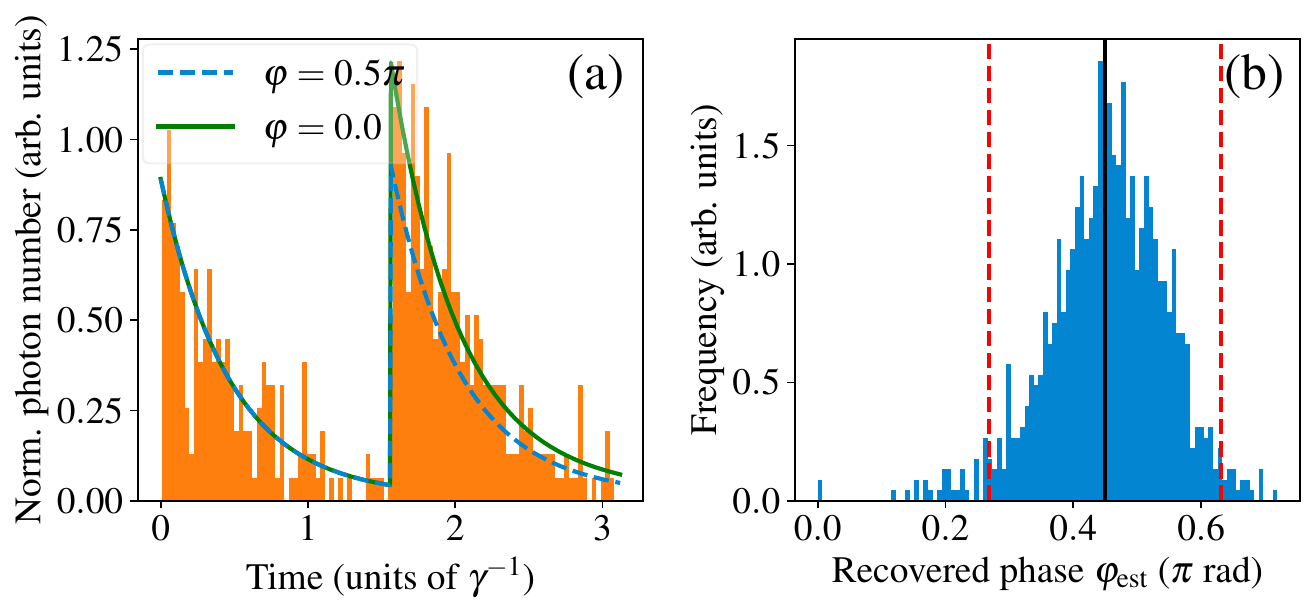}
\caption{Reconstruction of the relative pulse phase $\varphi$. (a) Time-dependent incoherently scattered intensities  for $\varphi = 0.5 \pi$ (green solid) and $\varphi = 0$ (blue dashed).  The orange histogram depicts one possible realization of a measurement with $N=500$ signal photons for $\varphi = 0$. (b) shows a histogram of recovered phases $\est$ obtained with $n=1000$ repetitions of the phase reconstruction. The true pulse-to-pulse phase $\varphi$ is indicated by the vertical black line. The red dashed lines indicate an interval symmetrically placed around $\varphi$ which contains 95~\% of the reconstructed phases. The target is a stainless-steel foil enriched in ${}^{57}$Fe with thickness parameter $b = 1.11 \gamma$. \label{fig-Stats}}
\end{figure}

To model the measurement outcome of a single experimental repetition with $N$ signal photons, we consider $I_{\mathrm{tot}}(t,T,\varphi)$ as a distribution function, and randomly draw arrival times of $N$ photons from it  using the statistical functions package \textsc{stats} of \textsc{scipy}~\cite{Scipy} in \textsc{python}~\cite{Python3}. Note that $N$ is the total number of scattered photons from both pulses. A first example with a thin target exhibiting near-exponential decay is given in Fig.~\ref{fig-Stats}(a), considering a stainless-steel foil ($\textrm{Fe}_{55}\textrm{Cr}_{25}\textrm{Ni}_{20}$) enriched to $95\%$ in ${}^{57}$Fe with optical thickness parameter $b = 1.11 \gamma$  corresponding to a physical thickness of $0.5~\mu$m. The pulse-to-pulse separation is given by $T=220~\mathrm{ns}\approx 1.6\,\gamma^{-1}$.
The solid green and dashed blue lines show the time-dependent incoherent intensities for two possible relative phases $\varphi = 0, \pi/2$, respectively. As expected, the initial decay after the first XFEL pulse is insensitive to $\varphi$. However, after the second pulse, the signal depends on $\varphi$, owing to interference in the scattering of the two pulses off the nuclei. This allows one to retrieve $\varphi$ provided that sufficient signal photons are measured. The orange histogram shows a single realization of $N=500$ signal photons and $\varphi = 0$.  From this histogram, an estimate $\est$ for the actual phase $\varphi$ is obtained by fitting $I_{\mathrm{tot}}(t,T,\varphi)$ with parameters $I_0$ and $\varphi$ using Poisson regression~\cite{PoissonMyers}. 

The recovered phase $\est$ fluctuates from realization to realization. To illustrate this, we repeat the above phase-reconstruction $n=1000$ times for a fixed $\varphi$, and form a histogram of the obtained $\est$ values. The result is shown in Fig.~\ref{fig-Stats}(b), with the true value $\varphi$  indicated as solid black line. We find that the distribution of estimated phase values $\est$ is well-centered around $\varphi$, showing that a successful reconstruction is indeed possible. Note that the number of realizations with  $\est = 0 $ is higher than expected. This is an artifact of the phase reconstruction, which favors the extremal  phases $\est = 0, \pi $ in particular for low photon numbers. We attribute this to a mutual dependence of the fit parameters $I_0$ and $\varphi$, which results in  an over- or underestimation of the contribution $\propto \cos(\varphi)$, if the  intensity $I_0$ cannot reliably be determined from a strongly fluctuating spectrum. This leads to an excess of phase recoveries in the edge cases $\est = 0, \pi$ which extremize the cosine function.

Next, we optimized  the target thickness parameter $b$ (see Appendix~\ref{app:target}). We found that an optimum value of $b_{\mathrm{opt}}$ exists, because the target thickness influences the residual intensity of the x-rays scattered from the first pulse after the arrival time of the second pulse. This residual scattering crucially determines the visibility of the interference signal required for determining $\varphi$. For incoherent scattering from single-line stainless steel targets as considered above, we determined $b_{\mathrm{opt}} = 2.68\gamma$, which corresponds to a physical thickness of  $1.2$~$\mu$m.

Finally, we turn to the systematic analysis of the  phase recovery based on the nuclear-incoherently scattered light. In particular, we aim at characterizing which information on $\varphi$ can be gained from a phase recovery result $\est$.  To this end, we repeated the above-described phase reconstruction for $N=1000$ photons and a stainless-steel target with  optimum target thickness $b_{\mathrm{opt}}$ in a range of pulse-to-pulse phases $\varphi \in [0,\pi]$. In all cases, we chose $n = 1000$ repetitions to reduce statistical fluctuations. From this data, we determine the distribution of true phase values $\varphi$ which result in a recovered phase $\est$. This distribution characterizes the information gained by measuring a particular phase $\est$: If the distribution is narrow, then $\est$ is likely to provide a good estimate of the actual phase $\varphi$. More precisely, for each phase $\varphi$, we determined the number of repetitions which resulted in a recovered phase in the interval $[\est-\delta \varphi/2,\est+\delta\varphi/2]$, where $\delta \varphi = \pi / 100$ is the step size of the $\varphi$ sampling. To allow for a comparison of the different $\est$-intervals, we then normalized each distribution to the total number of repetitions in the interval, summed over all $\varphi$ values.

\begin{figure}[t]
\includegraphics[width = \columnwidth]{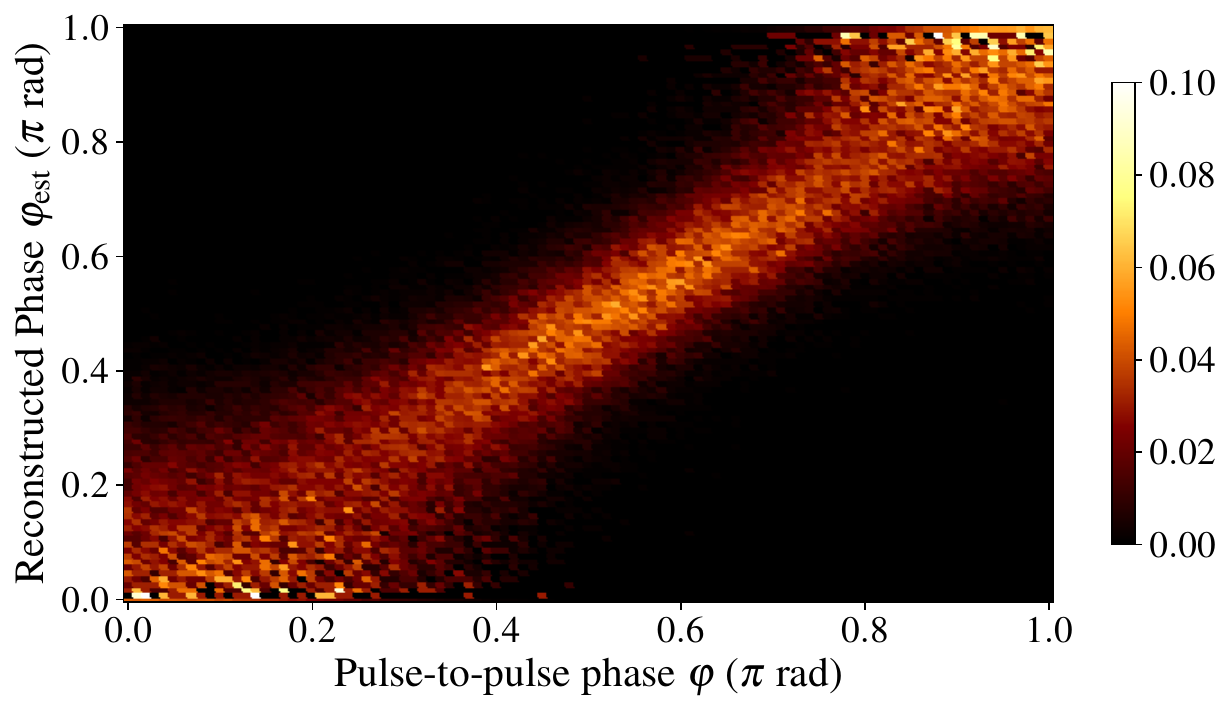}
\caption{Normalized distribution of $\varphi$ values resulting in a reconstruction result $\est$. The width of the distribution characterizes how well a reconstructed $\est$ constrains the range of possible  true phase values $\varphi$. Details are given in the main text. Results are shown for $N=1000$ signal photons. \label{fig-EstPlot}}
\end{figure}

The result is shown in Fig.~\ref{fig-EstPlot}. It can be seen that the reconstructed phase $\est$ is well-centered around the true phase $\varphi$. The width of the $\varphi$-distribution for a given $\est$ is most narrow in a region  around   $\est = \pi/2$. In this region, the true phase can be estimated best with high probability from  $\est$. This is consistent with the $\cos(\varphi)$-dependence of the interference contribution in Eq.~(\ref{eq-TwoLevelDist}), for which the  gradient of the intensity along $\varphi$ is highest  around $\varphi = \pi/2$, facilitating a reliable fit. Towards $\est = 0, \pi$ the distributions become broader, but still allows one to gain information on the most likely $\varphi$.

In Appendix~\ref{sec-PhotonScan}, we further characterize the phase recovery performance as function of the photon number $N \in [100,1050]$. As expected, the performance of phase reconstruction increases with  photon number, and provides reasonable accuracy above $N \approx 500$. We further found that the phase reconstruction fits for $\varphi$ close to $0, \pi$ tend to spuriously converge to $\varphi_\mathrm{est} = 0, \pi$,  in particular for low signal photon numbers, resulting in a lower success probability for the phase recovery  in that region of parameter space. This further supports the above result that the phase recovery works best around $\varphi = \pi/2$.

In summary, we have shown that long-lived M\"ossbauer nuclear resonances allow one to determine the relative phase $\varphi$ between two consecutive transform-limited XFEL-pulses. The phase recovery relies on Ramsey-like interference in the time-dependent intensity of the x-rays of both pulses scattered off the nuclear target. The  phase information can be used as additional input to a downstream primary experiment. Alternatively, it might be used as feedback to the XFEL source, to stabilize the pulse-to-pulse phase to a particular value. 
The measurement scheme crucially relies on the long lifetime of the nuclear resonances, which allows one to induce interferences across the temporal separation of the two XFEL pulses.   
We found that the phase recovery works best in a certain window of relative phases around $\varphi = \pi/2$. However, for future applications in coherent multi-pulse experimental schemes, other pulse-to-pulse phases may also be desirable. To this end, the optimum operation range of our approach can be shifted to other phase values, by inducing additional contributions to the interferometrically measured phase. For example, Doppler motion of the M{\"o}ssbauer sample, a technique widely used in M{\"o}ssbauer science~\cite{RoehlsbergerBook,coussement1996,coussement2000,PoundRebka}, allows one to detune the nuclear transition frequency by $\Delta = \nu-\omega_0$, giving rise to a displaced  $\cos(\varphi - \Delta T)$-dependence of the interference contribution  (see Appendix~\ref{app:shift}).

We envision that the phase-recovery method introduced here will open up new avenues towards accessing and coherently controlling dynamics on nano- and microsecond time scales currently not accessible  at accelerator-based x-ray sources. This also involves studies in which the sample dynamics between the two x-ray pulses is further modified by external stimuli. Possible physics platforms naturally include dynamics of biological, nanophysics and soft matter systems~\cite{Lehmkuehler2020, Lehmkuehler2021} and condensed matter systems probed via excitation of M{\"o}ssbauer resonances subject to external perturbations~\cite{bocklage2021,sadashivaiah2020, sadashivaiah2021, meyer2021,vagizov2007,Heeg2021}. However, it is important to note that M\"ossbauer nuclei are only used to gain information on the relative phase $\varphi$ of two x-ray pulses. In contrast, the primary experiment using these pulses may not involve nuclear resonances at all. Also, the phase-reconstruction approach presented here could also be generalized to other long-lived resonances operating at different frequency ranges or XFEL pulse separations $T$.

\begin{acknowledgments}
The authors would like to thank Miriam Gerharz 
and Deniz Ad{\i}g{\"u}zel for fruitful discussions.
\end{acknowledgments}

\appendix

\section{Dependence of the phase recovery on the signal photon number $N$ \label{sec-PhotonScan}}

In order to analyze the dependence of the phase recovery on the signal photon number $N$, we require a measure to quantify its performance. For this, we employ the half-width $\Delta \est$ of the distribution of $\est$ values sampled for a fixed pulse-to-pulse phase $\varphi$, as introduced in Fig.~\ref{fig-Stats}. It characterizes the range of possible $\est$ values which occur in the recovery for a given true phase $\varphi$. A more narrow width indicates a better correlation of $\varphi$ and the possible $\est$ values, implying a more accurate phase recovery on average.

The distribution is determined as described in the main text: The drawing of random samples and fitting of the phase $\est$ to the distribution Eq.~(\ref{eq-TwoLevelDist}) as illustrated in Fig.~\ref{fig-Stats}(a) is repeated $n = 1000$ times, to form histograms of recovered phases $\est$. Using such a  histogram, the range of possible $\est$ values for a given true phase $\varphi$ is expressed via the symmetric interval $[\varphi - \Delta \est, \varphi + \Delta \est]$ which contains $95 \%$ of the reconstructed values $\est$ around $\varphi$. One example of this is shown in panel (b) of Fig.~\ref{fig-Stats}. The interval ranges $\varphi \pm \Delta \est$ are indicated as red dashed lines, while the true value $\varphi$ is indicated as black solid line. This procedure is then repeated for signal photon numbers in the interval $N \in [100,1050]$ and phases in the interval $\varphi \in [0,\pi]$.

\begin{figure}[t]
\includegraphics[width = \columnwidth]{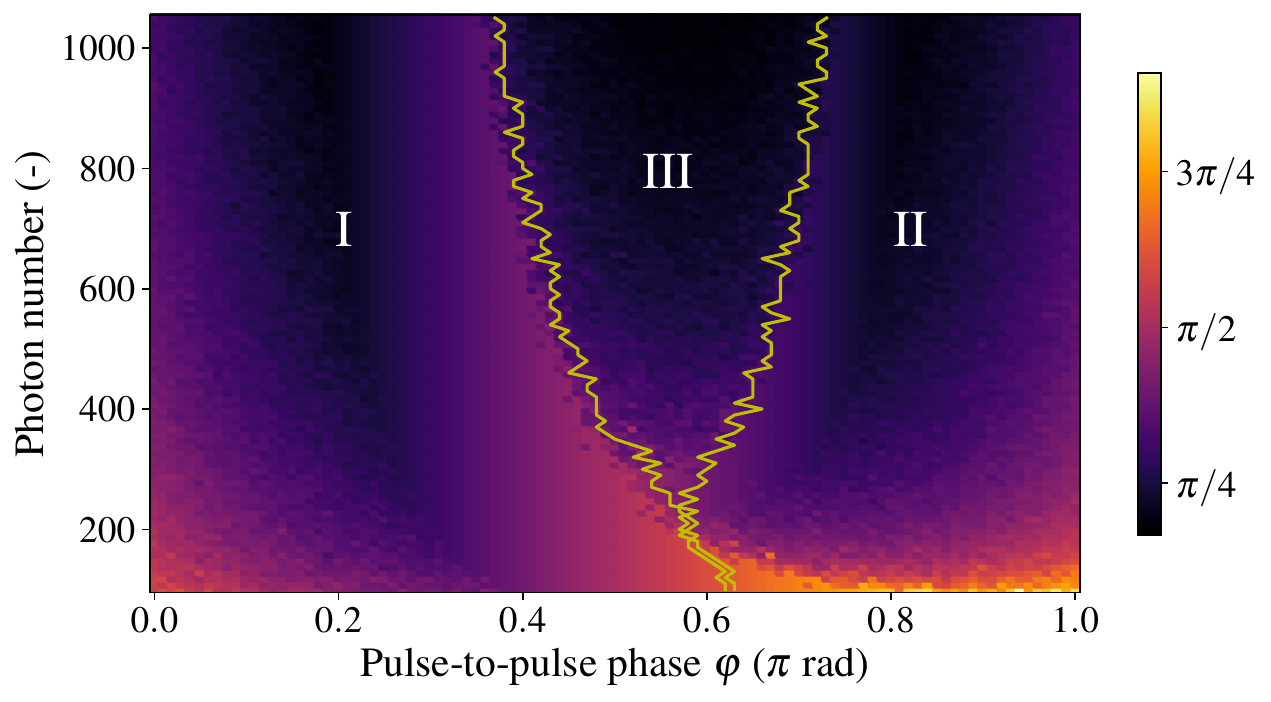}
\caption{Performance of the phase recovery as function of signal photon number $N$ and pulse-to-pulse phase $\varphi$. The color code shows the half-width $\Delta \est$ around the true value $\varphi$  which contains 95 \% of the sampled $\est$ values. The yellow curves indicate $\varphi$ values below (region I) or above (region II) which the number of reconstructions $ \est = 0, \pi$ exceeds any other reconstructed value in the interval $[0,\pi]$. In between (region III) lies the region with best performance.\label{fig-2DMap}}
\end{figure}

The result is shown in Fig.~\ref{fig-2DMap}, where $\Delta \est$ is depicted in color-coding as function of the pulse-to-pulse phase $\varphi$ and the number of signal photons $N$. As expected, the phase recovery generally improves with increasing signal photon number $N$, since then the individual spectra contain more information on  $\varphi$. However, interestingly, there is also a pronounced dependence of the recovery performance on the actual phase $\varphi$.  As expected from the $\cos(\varphi)$-dependence of the interference signature in Eq.~(\ref{eq-TwoLevelDist}) in the main text, $\Delta \est$ is comparably large close to $\varphi = 0, \pi$. Furthermore, the number of fits which spuriously converge to $\est = 0, \pi$ increases if the actual value $\varphi$ is close to these values. In Fig.~\ref{fig-2DMap}, in region I bounded by the yellow curve, the majority of the repetitions yielded $\est = 0$. Analogously, the majority of the repetitions resulted in $\est = \pi$ in region II. As a result, we find that the phase recovery clearly works best in region III, close to $\varphi = \pi/2$, and above $N\approx 500$.

\begin{figure}[t]
\includegraphics[width = \columnwidth]{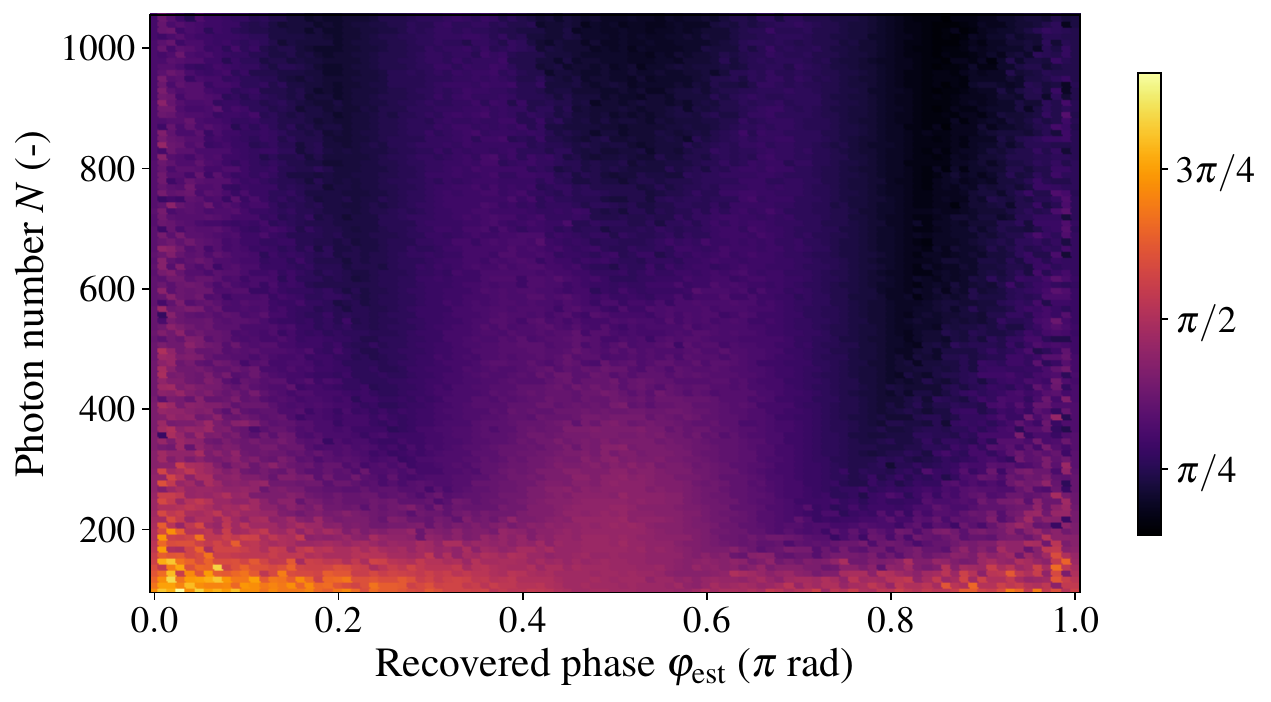}
\caption{Performance of the phase recovery as function of signal photon number $N$ and the recovered phase $\est$.  The color code shows the half-width $\Delta \varphi$ of the distribution of possible $\varphi$ values which may result in a given $\est$. The width is defined such that it comprises 95\% of the possible $\varphi$ values.\label{fig-EstScan}}
\end{figure}

Analogously, one may also analyze the distribution $\Delta \varphi$ of $\varphi$ values which result in a particular $\est$ value. This generalizes the result of Fig.~\ref{fig-EstPlot} to different signal photon numbers $N$, using the same computational approach as described in the main text. This distribution more directly quantifies how much information on the true phase $\varphi$ is gained from a particular recovered phase $\est$. The result is shown in Fig.~\ref{fig-EstScan}. The color code indicates the width $\Delta \varphi$ as a function of signal photon number $N$ and estimated phase $\est$. As expected, the width $\Delta \varphi$ decreases with increasing photon number $N$ leading to a more accurate correspondence between retrieved phase $\est$ and true phase $\varphi$. Close to the edge cases $\est = 0, \pi$ the reconstruction becomes worse owing to the high number of spuriously converging $\est$ values close to $\varphi = 0, \pi$. The most accurate results are found around values of $\est  = \pi/2$ for larger photon numbers. In the intermediate region the width $\Delta \varphi$ increases towards the center of the plot while towards the edge cases smaller widths $\Delta \varphi$ are found. This is caused by the fact that $\Delta \varphi$ values close to the edge cases $\varphi = 0,\pi$ are bounded by the neighboring edge from above.


Overall, we conclude that the phase recovery works best for estimated phases around $\est = \pi/2$ and photon numbers above $N \approx 500$, consistent with our observations in Fig.~\ref{fig-2DMap}. The similarity of the distribution widths in Figs.~\ref{fig-2DMap} and \ref{fig-EstScan}, especially for large photon numbers, is in good agreement with the observation of an equal width of the phase distribution along the $\varphi$ and $\est$ axes in Fig.~\ref{fig-EstPlot}. This establishes the recovered phases $\est$ as reliable estimates of the corresponding true phase $\varphi$.

\begin{figure}[t]
\includegraphics[width = 0.9\columnwidth]{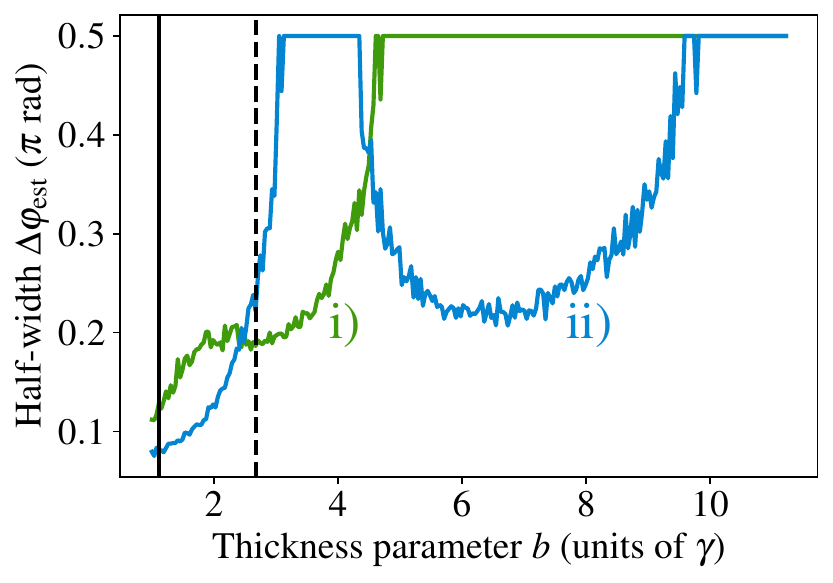}
\caption{Performance of the phase recovery as function of target thickness $b$. The figure shows  $\Delta \est$ evaluated for $\varphi = \pi/2$ with $N=1000$, which quantifies how closely the true phase $\varphi$ and the possible reconstruction outcomes $\est$ are correlated. Results are compared for (i) spatially incoherent scattering from a single-line stainless steel target, as considered in the main text, and (ii) enriched $\alpha$-iron with hyperfine splitting in nuclear forward scattering geometry. The vertical solid black line at $b = 1.11 \gamma$ indicates the thickness parameter used in Fig.~\ref{fig-Stats}. The vertical black dashed line corresponds to $b = 2.68 \gamma$, employed in Figs.~\ref{fig-EstPlot}-\ref{fig-EstScan}.\label{fig-Thickness}}
\end{figure}

\section{\label{app:target}Nuclear target optimization}

The performance of the phase reconstruction crucially depends on the choice of the nuclear target, in particular, on its thickness. The thickness not only influences the number of signal photons. Thicker targets also exhibit additional features such as non-exponential decay, dynamical beats, and superradiant acceleration of the initial decay dynamics~\cite{RoehlsbergerBook, Smirnov1999, HannonTrammell1999, kagan1999,vanBrck1999}. With hyperfine splitting, the time-dependent intensity further exhibits more rapid oscillations, the so-called quantum beats, which originate from the interference between scattering off different hyperfine transitions. These phenomena influence the visibility of the interference contribution which we use to recover the XFEL pulse-to-pulse phase $\varphi$.

In order to quantify the performance, we employ the half-width $\Delta \est$ of the distribution of sampled $\est$ values around the true pulse-to-pulse phase $\varphi$ as introduced in Fig.~\ref{fig-Stats}(b). In this figure, the distribution width $\Delta \est$ is indicated by the red dashed lines. For the analysis of $\Delta \est$, we choose the optimum working point $\varphi = \pi/2$ and $N=1000$. A small width implies that the true phase $\varphi$ is closely correlated to the reconstructed phase $\est$, i.e., on average a more accurate reconstruction.

The green curve (i) in Fig.~\ref{fig-Thickness} shows the dependence of $\Delta \est$ on the thickness parameter $b$ for a stainless-steel target without hyperfine splitting as considered in the main text. For comparison, the blue curve (ii) shows corresponding results for an $\alpha$-iron target enriched in ${}^{57}$Fe in nuclear forward scattering geometry. The magnetization is aligned  such that the two linearly polarized hyperfine transitions are driven by the XFEL pulses, leading to quantum beats in the time-dependent intensity. The respective time-dependent intensities used for the analysis are derived in Apendix~\ref{sec-Forward}.

For  thin targets, the half-widths $\Delta \est$ of the $\est$ distributions are narrow. With increasing thickness parameter $b$, the widths generally becomes larger, until they saturate towards $\Delta \est = \pi/2$ for higher $b$ values. At this saturation width, no useful information on the true phase $\varphi$ can be gained from a given $\est$. We attribute the saturation to the thickness effects, in particular the speed-up of the initial decay, as mentioned above. 

Interestingly, $\Delta \est$ does not monotonously increase with $b$, but assumes  local minima. For the forward-scattering case (ii), these structures are related to  the characteristic dynamical beat pattern, which affects the scattered light intensity resulting from the first XFEL pulse as function of time. If the nodes of the dynamical beats lie at times shortly after the arrival of the second XFEL pulse arrival, then the interference term $\propto \cos(\varphi)$ in Eq.~(2) of the main text is strongly suppressed and an accurate phase reconstruction becomes challenging. For the incoherent scattering case (i), the local minima likely arise from a similar mechanism, however, less direct due to the additional integration over the position $z$ in the sample. 

For the analysis of the phase recovery performance  in Fig.~\ref{fig-EstPlot}, we chose the local minimum of (i) at around $b = 2.68 \gamma$, indicated by the black-dashed line.  For the nuclear forward scattering case (ii), a corresponding local minimum appears at $b \approx 6.2 \gamma$. Note that we did not consider very thin targets in the main analysis even though they have smaller $\Delta \est$, since they are challenging to fabricate and lead to lower signal photon numbers. However, we show an example for $b = 1.11 \gamma$, corresponding to a $0.5\mu$m thick stainless-steel target, in Fig.~\ref{fig-Stats}. This value is indicated as vertical solid black line in Fig.~\ref{fig-Thickness}.

\section{\label{app:detuning}Derivation of the spatially incoherently scattered intensity with detuning}

Here, we derive the intensity of the x-rays spatially incoherently scattered off a nuclear sample of total thickness $z_{\textrm{max}}$, including a possible detuning between x-rays and nuclei which can be used to shift the optimum operation phase $\varphi = \pi/2$ to other values.

To this end, we first consider the field $E_{4\pi}(t,z)$ emitted by a  slice of  infinitesimal thickness at depth $z$ in the target. In linear response theory~\cite{Smirnov1999,Smirnov2007,HannonTrammell1999,RoehlsbergerBook},
\begin{align}
E_{4\pi}(t,z) = \int^{\infty}_{-\infty}dt'f_c(t-t')E_{\textrm{in}}(t',z) \,,\label{eq-LinResp}
\end{align}
where $f_c(t)$ denotes the single-nucleus dipole response~\cite{Smirnov1999}
\begin{align}
f_c(t) \sim -\frac{i}{2}e^{-i\omega_0 t}e^{-\frac{\gamma}{2}t}\Theta(t) \, , \label{eq-SingNuc}
\end{align}
with nuclear resonance frequency $\omega_0$ and single-nucleus line-width $\gamma$.
Note that we use ``proportional to'' here since we focus on the time-dependence of the incoherently scattered light, which is given by the absolute square of Eq.~(\ref{eq-LinResp}).  The magnitude depends on prefactors relating dipole response and field amplitude, as well as various experimental factors such as the detection geometry, and the branching ratios between different scattering processes such as forward scattering, incoherent $4\pi$ emission, scattering with or without recoil, or internal conversion~\cite{Smirnov2007,Bergmann1994,Sturhahn1995,sturhahn1999}.

The field $E_{\textrm{in}}(t,z)$ incident on the nuclei at depth $z$ in the nuclear target is given by
\begin{align}
E_{\textrm{in}}(t,z) = \int^{\infty}_{-\infty}dt'R(t-t',z)   E_{\textrm{in}}(t', z=0) \,, \label{eq:fwint}
\end{align}
i.e., the convolution of the field $E_{\textrm{in}}(t, z=0)$ incident on the target with the response function~\cite{Smirnov1999,HannonTrammell1999,RoehlsbergerBook}
\begin{align}
R(t,z) &= e^{-\frac{\mu_e}{2}z}\left[\delta(t) + S(t,z)\right] , \label{eq-NucResponse}\\[2ex]
S(t,z) &= -\sqrt{\frac{b(z)}{t}}\:e^{-\frac{\gamma}{2}t}\:e^{-i\omega_0 t}\:J_1\left (2\sqrt{b(z)t}\right)\Theta(t) \,.\label{eq-Bessel}
\end{align}
Here, $\mu_e$ denotes the electronic absorption length and $J_n$ the Bessel function of the first kind and order $n$. $b(z)$ is the optical thickness parameter up to depth $z$ along the light propagation axis defined as~\cite{RoehlsbergerBook}

\begin{align}
b(z) = \frac{1}{4}\rho \sigma_0 f_{\textrm{LM}}z\:\gamma\,,
\end{align}
where $\rho$ is the number density of the resonant isotope, $\sigma_0$ the nuclear resonance absorption cross section and $f_{\textrm{LM}}$ the Lamb-M{\"o}ssbauer factor.

The full response $R(t,z)$ comprises two contributions: The essentially unscattered direct ultrashort x-ray pulse described by the Dirac delta function and the delayed response $S(t,z)$ scattered off the upstream nuclei.

For a generic single incident XFEL pulse arriving at time $t=T$ with complex-valued amplitude $E_0$,
\begin{align}
    E_{\textrm{in}}(t, T, z=0) = E_0 \,e^{-i\nu t}\: \delta(t-T)\,,
\end{align}
we obtain from  Eq.~(\ref{eq:fwint})
\begin{align}
   & E_{\textrm{in}}(t, T, z) = E_0 \, e^{-i\nu T}  R(t-T,z)\,.
\end{align}
Inserting this field propagated to depth $z$ into Eq.~(\ref{eq-LinResp}), we find~\cite{Smirnov1999,SmirnovKohn1995,Smirnov2007,Shvydko1999,Adiguzel2024}
\begin{align}
    E_{4\pi}(t, T, z) \sim& -\frac{i\,E_0 }{2}\: e^{-\frac{\mu_e}{2}z}\:e^{-i\nu T}\:e^{-\left(i\omega_0 + \frac{\gamma}{2}\right)(t-T)}\times \nonumber \\[2ex]
    &\times J_0\left (2\sqrt{b(z)(t-T)} \right)\Theta(t-T) \,. \label{eq:e4pi}
\end{align}
This result characterizes the induced nuclear dipole excitation of an infinitesimal slice at depth $z$ sourced by the upstream x-ray light.
In the low-excitation limit, its absolute value squared relates to the spatially incoherently emitted intensity.

For the sequence of two consecutive XFEL pulses in Eq.~(1) of the main text,
\begin{align}
E_{\textrm{in}}(t, T, \varphi, z = 0) = E_0 \,e^{-i\nu t}\:\left[\delta(t) + e^{i\varphi}\delta(t-T)\right]\,,
\end{align}
we analogously find
\begin{align}
   E_{4\pi}(t,  T&, \varphi, z) \sim -\frac{i\,E_0 }{2}\: e^{-i\omega_0 t} \biggl [ \mathcal{E}_{4\pi}(t,z)\Theta(t)  \nonumber\\[2ex]
   &  + e^{i(\varphi - \Delta T)} \mathcal{E}_{4\pi}(t-T,z)\Theta(t-T) \biggr]\,. \label{eq:field-double-pulse}
\end{align}
Here, for notational brevity, we have defined the real-valued envelope of the response function
\begin{align}
    \mathcal{E}_{4\pi}(t,z) =  e^{-\frac{\mu_e}{2}z} \: e^{-  \frac{\gamma}{2}t}\:  J_0\left (2\sqrt{b(z)t} \right)\,, \label{eq:e-env}
\end{align}
also given as Eq.~(3) in the main text, and $\Delta = \nu - \omega_0$ is the detuning between x-ray carrier frequency  and nuclear resonance frequency.

From Eq.~(\ref{eq:field-double-pulse}), the emitted field intensity per slice evaluates to
\begin{align}
&I_{4\pi}(t, T, \varphi, z) =I_0\Theta(t) \biggl [ \mathcal{E}^2_{4\pi}(t,z) +
\mathcal{E}^2_{4\pi}(t-T,z)\Theta(t-T)  \nonumber \\[1ex]
&  + 2\cos(\varphi-\Delta T) \mathcal{E}_{4\pi}(t,z)\mathcal{E}_{4\pi}(t-T,z)\Theta(t-T) \biggr ] \,, \label{eq:i-slice}
\end{align}
in which the prefactors are combined into the overall intensity $I_0$. In the resonant case $\Delta = 0$, this reduces to Eq.~(2) in the main text.

Finally, to obtain the total signal $I_{\mathrm{tot}}(t,T,\varphi)$ of the entire sample, we integrate the intensity Eq.~(\ref{eq:i-slice}) emitted by a slice at position $z$ over the whole sample thickness from $z = 0$ to $z =z_{\textrm{max}}$,
\begin{align}
I_{\textrm{tot}}(t,T,\varphi) = \int^{z_{\textrm{max}}}_{0}I_{4\pi}(t, T, \varphi, z)\:dz \,.
\end{align}

\section{\label{app:shift}Shifting of the optimum phase recovery working point via the detuning}

In the main text, we found that the phase recovery performs best around the relative phase $\varphi = \pi/2$ (or equivalent phase ranges shifted by multiples of $\pi$). One reason is that around this phase, the dependence of $\cos(\varphi)$ on $\varphi$ is highest, facilitating the fit. Further, in this phase range  the number  of retrieved phase values is minimized which spuriously converge  to the edge cases $\est = 0, \pi$ at low photon number $N$.

However, the optimum range was determined assuming resonant interaction between the nuclei and the XFEL pulses.
In case of a detuning $\Delta$ between the XFEL carrier frequency and the nuclear resonance frequency, from Eq.~(\ref{eq:i-slice}) we find that the dependence of the interference contribution on $\varphi$ generalizes to
\begin{align}
    \cos(\varphi-\Delta T)\,. \label{eq:shift}
\end{align}
As a result, we find that then the optimum phase recovery range can be shifted to other $\varphi$ values via a detuning $\Delta$. Such detunings can be controlled, e.g., by linear sample motion with velocity $v$ inducing a Doppler shift $\Delta =\omega_0\,v/c$ relative to the nuclear resonance frequency $\omega_0$ at rest, which is a well-established technique in M{\"o}ssbauer science~\cite{RoehlsbergerBook,coussement1996,coussement2000,PoundRebka}.

\section{Phase recovery via coherent nuclear forward scattering \label{sec-Forward}}

In the main text, the presentation focused on incoherent nuclear scattering. Here, we show that also coherent nuclear forward scattering (NFS) can analogouly be used to retrieve the phase between two XFEL pulses.

For a single-line  nuclear target with thickness parameter $b$, the NFS field is given by Eqs.~(\ref{eq:fwint}-\ref{eq-Bessel}). With the double-pulse defined in Eq.~(1) in the main text, the propagated field evalutes to
\begin{align}
E_{\mathrm{NFS}} &= E_0 \: e^{-i\omega_0 t}\biggl [ \mathcal{S}(t) + e^{i(\varphi - \Delta T)}\mathcal{S}(t-T) \biggr ]\,, \label{eq:single-field} \\[2ex]
\mathcal{S}(t) &= -e^{-\frac{\mu_e}{2}z}\: \sqrt{\frac{b}{t}}\:e^{-\frac{\gamma}{2}t} \:J_1\left (2\sqrt{b t}\right)\,, \label{eq:s-envelope}
\end{align}
where $\mathcal{S}(t)$ is the real-valued envelope of the delayed contribution of the response function, defined in analogy to Eq.~(\ref{eq:e-env}). Note that  we have omitted the ``prompt'' contributions related to the $\delta(t)$-part of the response function Eq.~(\ref{eq-NucResponse}). It comprises the enormous off-resonant background of the incident XFEL pulses and usually cannot be resolved in the experiment due to detection limitations.
From Eq.~(\ref{eq:single-field}), the NFS intensity evaluates to
\begin{align}
I_{\mathrm{NFS}}&(t,T,\varphi) = I_0 \: \Theta(t) \biggl [    \mathcal{S}^2(t) +  \mathcal{S}^2(t-T) \Theta(t-T)   \nonumber\\[1ex]
& + 2\cos (\varphi-\Delta T)\, \mathcal{S}(t) \mathcal{S}(t-T) \Theta(t-T)\biggr ] \,.\label{eq-SingleInt}
\end{align}
We find that the structure of the intensity Eq.~(\ref{eq-SingleInt}) is in complete analogy to that of the intensity emitted incoherently by a single target slice Eq.~(\ref{eq:i-slice})  (a further integration over $z$ is not required since  Eq.~(\ref{eq-SingleInt}) already describes the total NFS field downstream the target). Most importantly, it contains the same dependence of the interference contribution on $\cos(\varphi - \Delta T)$ and therefore also provides access to the desired phase information.

Finally, we also consider the case with hyperfine splitting, i.e., with multiple nuclear resonances. These typically give rise to so-called quantum beats, i.e., comparably fast oscillating structures in the time-dependent intensity due to interference between different scattering channels within the individual nuclei~\cite{HannonTrammell1999,RoehlsbergerBook}. One motivation for this is that in experiments, the $\delta(t)$-like prompt contributions in forward direction can be suppressed using x-ray polarimetry~\cite{Toellner1995,Marx-Glowna:hf5410}, if the nuclear target is optically active and can rotate the x-ray polarization. If polarizer and analyzer are operated in crossed setting, then the prompt parts without polarization rotation can be blocked, while the nuclear scattering parts with rotated polarization may pass the analyzer. This approach necessarily requires a hyperfine splitting to achieve the optical activity.

As the simplest example, we consider two hyperfine resonances, symmetrically split from the bare resonance frequency $\omega_0$ by $\pm \beta$. In the   limit of well-separated resonances, the response function of this setting can be approximated by the sum of two individual response functions Eq.~(\ref{eq-NucResponse})~\cite{RoehlsbergerBook},
\begin{align}
S(t,z&,\delta) = \delta(t) e^{-\frac{\mu_e}{2}z} + \,\mathcal{S}(t)\:e^{-i(\omega_0 - \beta)t} \:\Theta(t) \nonumber \\[1ex]
&\qquad   \qquad + \,\mathcal{S}(t)\:e^{-i(\omega_0 + \beta)t} \:\Theta(t) \nonumber \\[2ex]
&= \delta(t) e^{-\frac{\mu_e}{2}z}  + 2 \,\mathcal{S}(t) \cos(\beta t)\:e^{-i\omega_0 t}\:\Theta(t)\,, \label{eq:s-approx}
\end{align}
where $\mathcal{S}(t)$ is defined in Eq.~(\ref{eq:s-envelope}) with appropriately chosen optical thickness parameter $b(z)$ taking into account the relative weights of the respective hyperfine transitions~\cite{RoehlsbergerBook}. Note that the linear superposition of the two individual response functions is only approximate, and assumes that the two lines are sufficiently detuned such that their spectra do not overlap.

From Eq.~(\ref{eq:s-approx}), we find that the hyperfine splitting implies the replacement
\begin{align}
     \mathcal{S}(t) \to  2 \,\mathcal{S}(t)  \cos(\beta t)\,.
\end{align}
The additional $\cos(\beta t)$-modulation of the time-dependent amplitude corresponds to the quantum beats. However, it does not change the general structure of the resulting intensity Eq.~(\ref{eq-SingleInt}), including the interference contribution. Therefore, we conclude that also in the presence of hyperfine splitting, the desired phase $\varphi$ between the XFEL pulses can be retrieved, if the additional quantum beat modulation is considered in the fit of the experimental data.

%

\end{document}